\documentclass[aps,twocolumn,prl,superscript,floatfix,superscriptaddress]{revtex4-2}

\usepackage{epsf}
\usepackage{epsfig}
\usepackage{graphicx}
\usepackage{dcolumn}
\usepackage{braket}
\usepackage{bm}
\usepackage{amsfonts}
\usepackage{amsmath}
\usepackage{amssymb}
\usepackage{color,soul}
\usepackage{wasysym}
\usepackage{mathrsfs}
\usepackage{natbib}
\usepackage{float}
\usepackage{multirow}

\usepackage[colorlinks=true,%
            linkcolor=blue,%
            urlcolor=blue,%
            citecolor=blue,%
            filecolor=blue,%
            bookmarksopen=true,%
            pdfauthor={M.Figueroa},%
            pdftitle={QR},%
            pdfsubject={QR_manuscript},%
            pdfpagemode=UseOutlines]{hyperref}

\usepackage[T1]{fontenc} 
\usepackage{lmodern}

\usepackage{cancel}
\usepackage{xcolor}
\usepackage{mathtools}

\begin{document}

\title{Bound states in the continuum in whispering gallery resonators with pointlike impurities}
\author{M.A. Figueroa}
\affiliation{Departamento de F\'isica, Universidad T\'ecnica Federico Santa Mar\'ia, Casilla 110, Valpara\'iso, Chile}

\author{Vladimir Juri\v ci\'c}\thanks{Corresponding author:vladimir.juricic@usm.cl}
\affiliation{Departamento de F\'isica, Universidad T\'ecnica Federico Santa Mar\'ia, Casilla 110, Valpara\'iso, Chile}

\affiliation{Nordita, KTH Royal Institute of Technology and Stockholm University, 
Hannes Alfvéns väg 12, SE-106 91 Stockholm, Sweden}
\author{P.A. Orellana}\thanks{Corresponding author:pedro.orellana@usm.cl}
\affiliation{Departamento de F\'isica, Universidad T\'ecnica Federico Santa Mar\'ia, Casilla 110, Valpara\'iso, Chile}

\begin{abstract}
{Whispering-gallery resonators offer a versatile platform for manipulating the photonic transmission. Here, we study such a system, including periodically distributed pointlike impurities along the resonator with ring geometry. Based on an exact expression for the transmission probability we obtain here, we demonstrate that the bound states in the continuum (BICs) form from the whispering gallery modes at the high-symmetry momenta in the ring's Brillouin zone. Furthermore,  the presence of the inversion symmetry allows for a selective decoupling of resonant states, favoring the BIC generation and, therefore, allowing extra tunability in the optical transmission of the system. }
\end{abstract}
\maketitle

\emph{Introduction.} Bound states in the continuum (BICs) remain localized and coexist with a continuous spectrum of radiating states that propagate outside the system \cite{BICreview,BICsShereena}. Although this special class of states was proposed at the dawn of quantum mechanics by von Neumann and Wigner \cite{vonNeumann}, only in the following decades did such states become the subject of more intense research \cite{BICsStillinger, Friedrich}. Furthermore, the concept of BICs is not exclusively operative in quantum mechanics but also pertains to classical wave phenomena, e.g., in photonic, acoustic, and electronic setups~\cite{TrappedLight,AcousticFano,AcousticBIC,GhostFano,GuevaraOrellana,BICPhotonic,Longhi2013,InteractingBIC,Zhang2023}. Due to advancements in nanofabrication technology and concomitant experimental observation of the BIC and BIC-like states in these systems~\cite{BICsObservation,Plotnik2011,Gansch2016,Azzam2018,NanoLetters2020,Yu2020,Kronowetter2023}, they have also recently attracted considerable interest for applications, as in building lasers, photonic circuit elements, sensors, and filters \cite{BICsOptic,Lasing,Laser,Zou2015,LiuNanoLetters2023,Biosensor,filter}. In different classes of BICs, the photonic ones are of great importance both fundamentally and for the applications mainly because they can confine light~\cite{BICreview}, and the main focus of this paper is on photonic BICs.

In this respect, whispering gallery resonators (WGRs) are of a special importance due to their unique properties that may facilitate BIC formation. For example, they can exhibit resonances with high-quality factors and small volumes, which makes them helpful in improving light-matter interactions~\cite{high-Q-2006}. Furthermore, these systems have diverse applications, such as optical switches \cite{WGR5}, routers \cite{WGR6}, and sensors for pathogens and nanoscale objects \cite{WGR2,WGR3,WGR4}. Ideally, a WGR exhibits a symmetric spectrum { due to the spatial inversion symmetry of the impurity configuration}, implying the degeneracy between clockwise and counterclockwise propagating modes. { When the system is coupled to a continuum of modes, this symmetry is generically broken, and the modes hybridize. But,  for a special,  inversion-symmetric configuration, the antisymmetric mode corresponds to the BIC~\cite{WGR1}, which becomes a quasi-BIC once this symmetry is broken.} Such symmetry breaking can be caused by built-in heterogeneities like surface or bulk imperfections when a WGR is coupled to a continuum \cite{WGR7}.
\begin{figure}[t]
    \centering
\includegraphics[width=\linewidth]{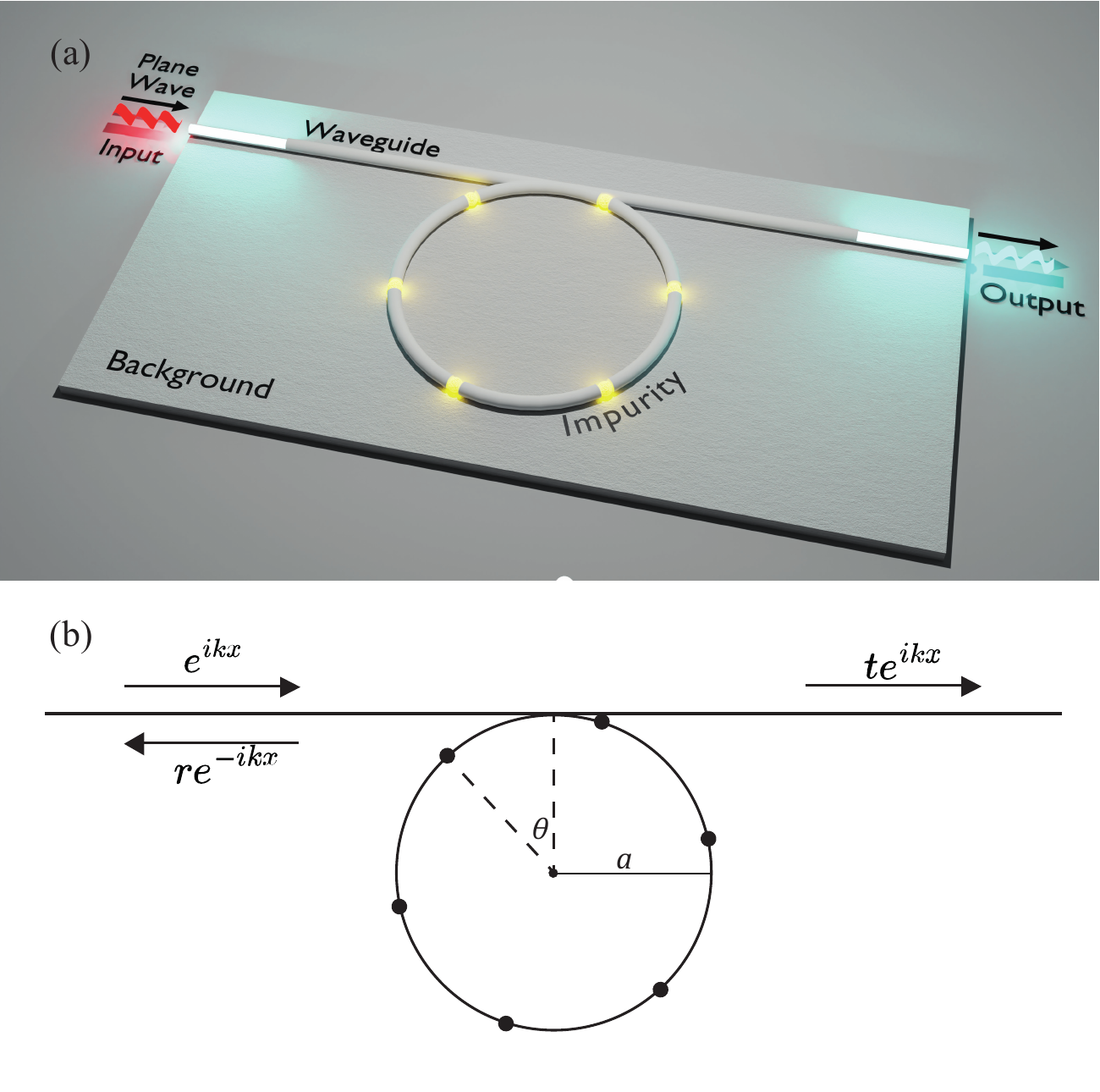}
    
    \caption{Schematic view of a whispering gallery resonator (WGR) with impurities coupled to the one-dimensional waveguide. (a) Illustration of the setup. (b) Details of the WGR of radius $a$ with $N$ point-like impurities (dark circles), separated by the angle $\Delta\phi=2\pi/N$, with the perturbation given by Eq.~\eqref{impurity}. The coupling angle ($\theta$) is defined as the angle between the first impurity on the left with respect to the line connecting the junction and the center of the WGR. Plane waves are incident from the left-hand side of the waveguide and are scattered by WGR.}
        \label{scheme}
\end{figure}

\begin{figure}[t!]
        \centering
        \includegraphics[width=\columnwidth]{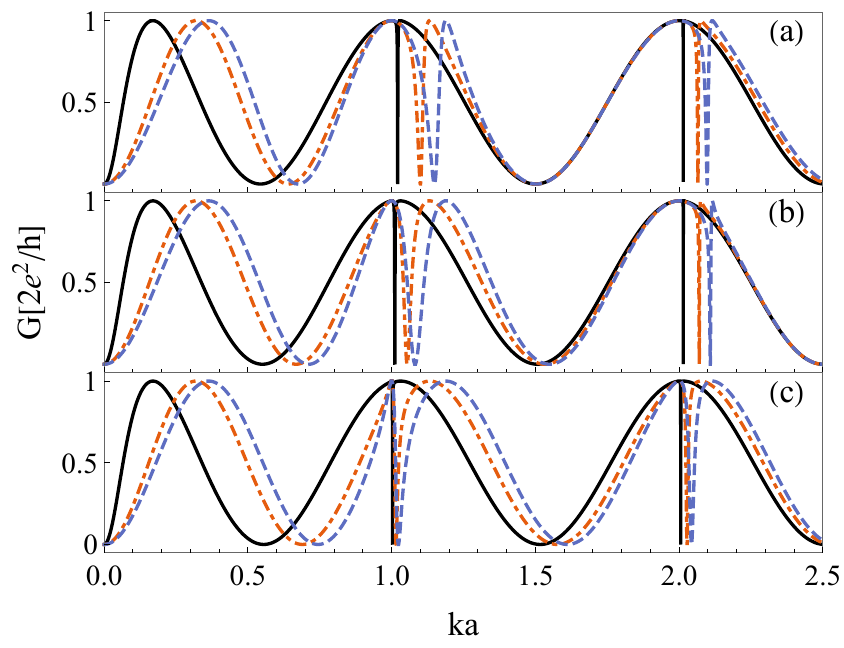}
        \caption{Dependence of the transmission $T$ on the wavenumber $ka$ for a tangential whispering gallery resonator with one impurity for different values of the coupling angle $\theta$ and the impurity potential, as given by Eq.~\eqref{transmission one imp}.  (a) $\theta=0.7\pi$, (b) $\theta=0.8\pi$ and (c) $\theta=0.9\pi$.  For a fixed $\theta$, the conductance is shown as the dimensionless impurity potential increases: $v_0 = 0.1$ (black solid line), $v_0 = 0.5$ (red dashed-dotted line), and $v_0=0.8$ (blue dashed line).  As $v_0$ increases, the resonant wavenumbers gradually modify and eventually split, leaving those with integer values of $ka$ unaltered. On the other hand, it can observed that the change in the coupling angle affects the wavenumbers of anti-resonances and the asymmetric Fano profiles.}
        \label{OneImpurity1}
\end{figure}

\begin{figure}[t!]
        \centering
        \includegraphics[width=\columnwidth]{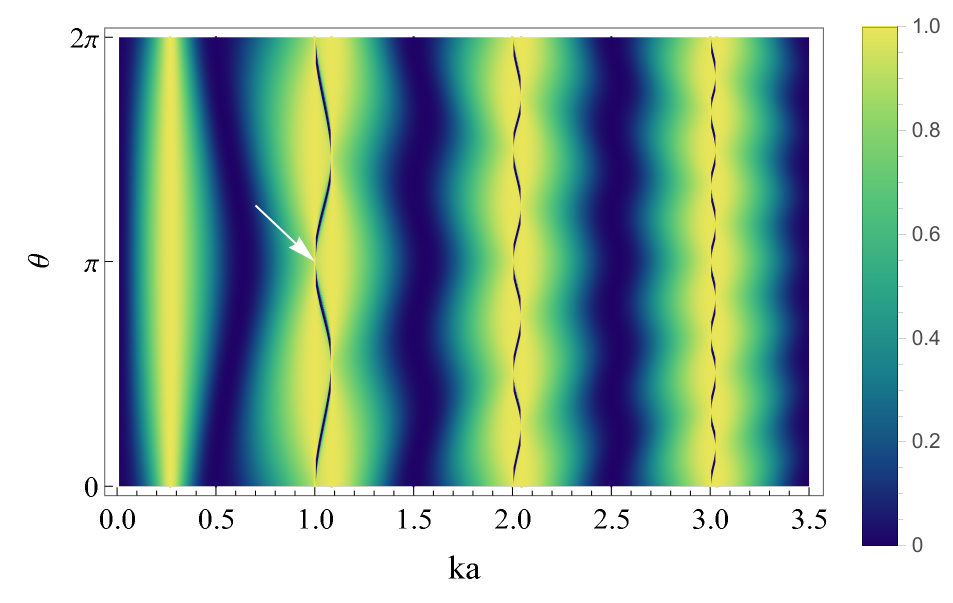}
        \caption{Transmission $T$  as a function of the coupling angle $\theta$ and the wavenumber $ka$ for a whispering gallery resonator with one impurity and the dimensionless coupling $v_0=0.3$ [see Eq.~\eqref{transmission amplitude}]. The white arrow indicates the collapse at the lowest energy. Notice that when the minima and maxima of the conductance coincide, the Fano profile collapses. Color code corresponds to the values of the transmission in the plot.}
        \label{OneImpurity2}
\end{figure}

\begin{figure}[t]
        \centering
\includegraphics[width=\columnwidth]{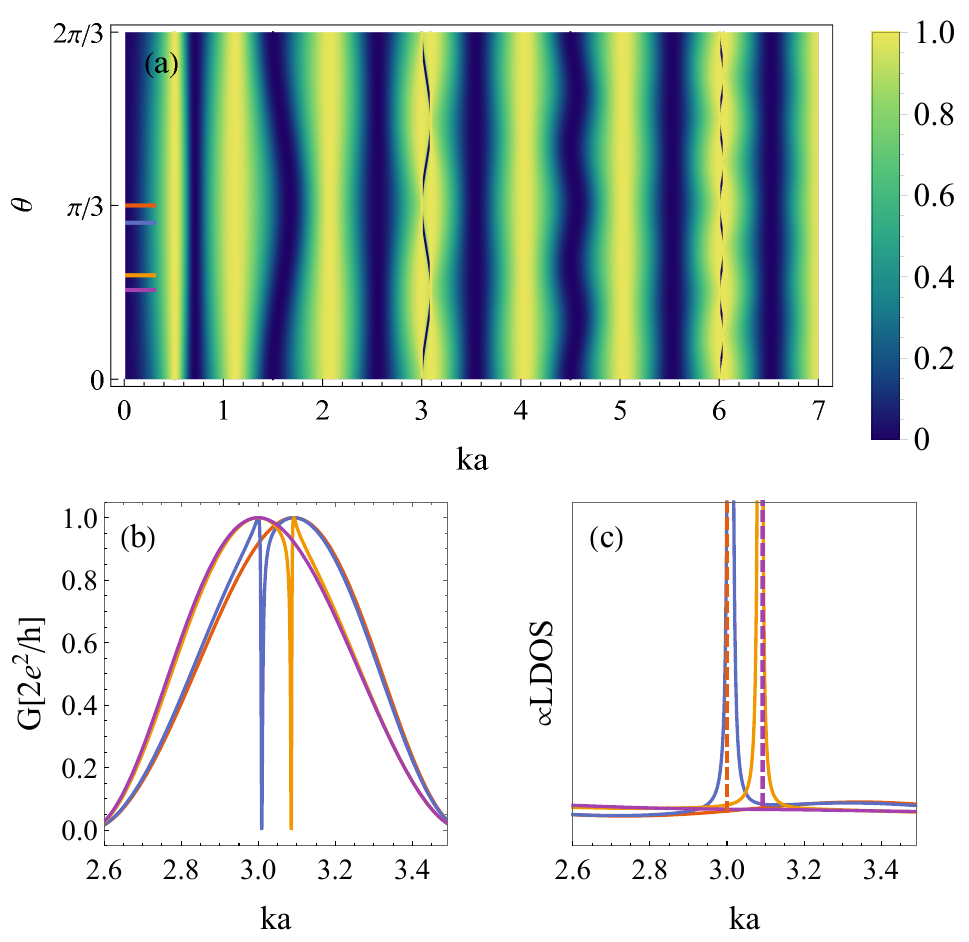}
        \caption{(a) Transmission ($T$) as a function of the coupling angle $\theta$ and the dimensionless wavenumber, $ka$, for a whispering gallery resonator with three impurities and the dimensionless impurity potential $v_0=0.3$. (b) Horizontal line-cuts for $\theta=0.5\Delta\phi$ (red line), $\theta=0.45\Delta\phi$ (blue line), $\theta=0.3\Delta\phi$ (yellow line) and $\theta=0.26\Delta\phi$ (purple line), obtained from Eq. \eqref{transmission amplitude}, with $\Delta\phi=2\pi/3$ as the angle between the nearest-neighbor impurities. (c) The local density of states for the same values of the parameters.}
        \label{N3 Listplot2}
\end{figure}

Motivated by these developments, in this work, we propose a novel route for the BIC manipulation in a WGR: The scattering by the pointlike impurities. In particular, we investigate the photonic transmission and BIC manipulation in a WGR coupled to a one-dimensional waveguide in conjunction with the effects of many impurities, as illustrated in Fig.~\ref{scheme}(a), with a detailed setup shown in Fig.~\ref{scheme}(b). The obtained
transmission probability is explicitly shown in Eq.~\eqref{transmission amplitude}. 
Based on this result, we subsequently prove that when there is only one impurity, the system can host BICs that arise from the collapse of the Fano profile in the transmission probability { when the system features the spatial inversion symmetry}. This occurs when the position of the impurity relative to the junction changes  (Fig.~\ref{OneImpurity2}), with the minima in the transmission explicitly shown in  Fig.~\ref{OneImpurity1}. The cause of this phenomenon is the destructive interference between the propagating states along the WGR and their subsequent scattering from the single impurity.
Furthermore, by considering the case of three impurities, we show that the BIC emerges { for the inversion-symmetric configuration}, as explicitly displayed in terms of the Fano profile collapse and the corresponding peak in the local density of states (Fig.~\ref{N3 Listplot2}). 
We finally discuss the general features of $N$ impurities and the possible relevance of this setup for manipulating BICs in photonic and electronic transport.

\emph{Model and Method.} We here consider a quasi-one-dimensional system, a narrow optical waveguide featuring a single optical mode with wavenumber $k$ coupled to a micro-ring resonator of radius $a$
and with $N$ symmetrically distributed identical point-like scatterers, with the local perturbation for the 
$l^{\rm th}$ scatterer of the form
\begin{equation}\label{impurity}
    V^{(l)}(x)= V_0\, \delta(x-a\phi_l),
\end{equation}
as illustrated  in Fig.~\ref{scheme}. Here,   $V_0$ is the perturbation strength, assumed to be equal for all the impurities, and $\phi_{l} = 2\pi l/N - \theta $ is the angular coordinate of the $l^{\rm th}$ impurity, $l=1,2,...,N$.  A special role is played by the angle $\theta$, defining the relative angle between the last impurity and the junction,  which we refer to as the coupling angle [Fig.~\ref{scheme}(b)].   Notice that particular distribution of impurities with  $\theta=\Delta\phi/2$ enjoys an additional inversion-like symmetry $\phi\to -\phi\,\, ({\rm mod}\,\, 2\pi)$, with $\Delta\phi=2\pi/N$ as the angle between the nearest-neighbor impurities, implying particular features in the optical transport, as discussed below.

We employ the standard formalism to address the optical transport in this setup~\cite{Ramy2007,ozdemir2019parity}, as detailed in Sec.~S1 of the Supplementary Material (SM). We find an exact expression for the transmission probability ($T$),  for any number of the impurities and  coupling angle (Sec.~S1 of the SM, which includes Ref.~\cite{Griffith}), which  reads as
\begin{equation}\label{transmission amplitude}
   T = \frac{ \left\{U_{N-1}[f(ka)] g(ka,\theta) \right\}^2}{\{1 - T_N[f(ka)]\}^2 + \left\{   U_{N-1}[f(ka)] g(ka,\theta) \right\}^2}. 
\end{equation}
Here, $T_N$ and $U_N$ are the order-$N$ Chebyshev polynomials of the first and second kind, respectively, while  the functions $f$ and $g$ are defined as 
\begin{align}\label{functions}
   f(ka) &= {\rm Re}\left\{\left( 1 -i\frac{v_0}{ka} \right) e^{ika\Delta\phi}\right\} \\
   g(ka,\theta) &= {\rm Im}\left\{\left( 1 -i\frac{v_0}{ka} \right) e^{ika\Delta\phi}\right\} \nonumber\\
&   + \frac{v_0}{ ka} \cos{[ka(\Delta\phi-2\theta)]}, 
\end{align}
where $v_0 = \varepsilon V_0 a /2$ is the dimensionless impurity potential, while $\varepsilon$ is the average dielectric constant of the medium. 
We now analyze the obtained transmission probability in the case of a single impurity. 



\emph{Case of a single impurity.}  The transmission probability [Eq. (\ref{transmission amplitude})] for a single impurity takes the form  
\begin{widetext}
\begin{align}\label{transmission one imp}
   T &= \frac{ \left\{  \sin{[2\pi ka]} - \frac{v_0}{ka}\cos{[2\pi ka]} + \frac{v_0}{ka} \cos{[ka\left( 2\pi - 2\theta\right)]}  \right\} ^2}{\left\{1 - \cos{[2\pi ka]} - \frac{v_0}{ka} \sin{[2\pi ka]}\right\}^2 +    \left\{  \sin{[2\pi ka]} - \frac{v_0}{ka}\cos{[2\pi ka]} + \frac{v_0}{ka} \cos[ka\left( 2\pi - 2\theta\right)  \right\}^2}. 
\end{align}
\end{widetext}
In Fig.~\ref{OneImpurity1}, we show the corresponding transmission coefficient as a function of the (dimensionless)  wavenumber $ka$ for different local potentials ($v_0$) and junction angle ($\theta$). As observed, the transmission coefficient exhibits an oscillating behavior with narrow Fano profiles. The WGR acts as a resonator and defines the natural frequencies (wavenumbers) that facilitate transmission. As shown in  Figs.~\ref{OneImpurity1}, the degeneracy between clockwise and counterclockwise propagating modes is lifted when a single impurity is included { since the spatial inversion symmetry is broken for a generic impurity configuration}. The Fano profiles emerging from these modes are highly dependent on the system parameters and are fundamentally related to partially confined leaky modes (quasi-BICs). This relation can be observed through the variation in the minima and the characteristic width with the coupling angle.

To get better insight into this phenomenon, we plot the transmission probability as a function of the dimensionless wavenumber $ka$ and the coupling angle  $\theta$ in Fig.~\ref{OneImpurity2}. As can be observed, the abrupt drop disappears in many configurations (often referred to as a ghost-Fano resonance~\cite{GhostFano} and indicated by the white arrow). The most notable cases include the inversion-symmetric configuration ($\theta \to \pi$) and when the impurity is placed exactly at the junction ($\theta \to 0$). Further support for our findings is displayed in Fig.~S1 of the SM. As we explicitly show, at points where the transmission coefficient exhibits the Fano profile, the local density of states (LDOS) displays a narrow profile that reduces its characteristic width when $\theta \to \pi$ { (as the system approaches the spatial-inversion-symmetric configuration)}.  This behavior implies that one of the ring states decouples from the delocalized incident waves on the waveguide. In turn, it becomes localized, forming a BIC in this limit. In the following, we analyze how these results generalize to an arbitrary number of impurities. 

\emph{Generalization to $N$ impurities.}  The situation involving several impurities shares many properties with the previous case. Due to the discrete translational symmetry of the impurity configuration, the natural discrete frequencies of the isolated WGR are given by the transcendental equation (Sec.~S3 of the SM):
\begin{equation}
\begin{aligned}\label{trascendental eq}
   \cos{(qa\Delta\phi)} = \cos{(ka\Delta\phi)} + \frac{v_0}{ka} \sin{(ka\Delta\phi)}\equiv f(ka),
\end{aligned}
\end{equation}
where the integer Bloch's momentum $qa$ is defined in the first Brillouin zone. Eq.~\eqref{trascendental eq} can be interpreted as a dispersion relation, which is identical to the spectrum in the one-dimensional Kronig-Penney model \cite{KronigPenney,Griffiths_Schroeter_2018}. We emphasize that the correspondence with the resonant modes is primarily due to the matching of the periodicity of the isolated WGR and the Griffith boundary condition~\cite{Griffith} applied at the coupling point. 

To illustrate the effects for several impurities, we plot in Fig.~\ref{N3 Listplot2}(a) the transmission probability as a function of the dimensionless wavenumber ($ka$) and $\theta$ for three impurities with a potential strength $v_0=0.3$. Analyzing the band structure, we find that Fano profiles specifically manifest within the band gap, i.e., between wavenumbers with high-symmetry Bloch momenta (at the center and edges) in the Brillouin zone. These WGR modes, with wavelengths comparable to the impurity spacing ($\lambda \sim a \Delta \phi/m$, $m \in \mathbb{N}$), interfere with waveguide modes due to their non-degeneracy. On the other hand, the interference governed by the continuity condition at the coupling point leads to the formation of BICs when the coupling approaches a node, i.e., at the point of destructive interference. This is illustrated by horizontal cuts for different coupling angles, as indicated by the colored lines in Fig.~\ref{N3 Listplot2}(a), which are displayed in Fig. \ref{N3 Listplot2}(b). Furthermore, Fig.~\ref{N3 Listplot2}(c) shows the LDOS  for the same parameters; see also Sec. S2 of the SM for additional details. The divergence and collapse of the delta-like profile of the LDOS further demonstrate the decoupling of the WGR state at coupling angles corresponding to nodes (Sec. S3 of the SM).

 Let us now analyze the transmission near the inversion-symmetric impurity configuration where a BIC can be formed, i.e., in the neighborhood of $\tilde{x}_m \equiv k_ma=mN$ ($m \in \mathbb{N}$), with respect to the potential strength and coupling orientation. As a result, the Fano profile is obtained by expanding the transmission coefficient in a series. Defining $\theta = \Delta\phi/2-\Delta/2$, $\tilde{x}=ka$, and keeping terms such that ${\rm max}(|\tilde{x}-\tilde{x}_m|) \sim \Delta^2$, the corresponding  transmission probability takes the form  
\begin{equation}
    T\approx \left(\frac{1}{1+q_F^2}\right)
    \frac{\left(\tilde{X} - q_F \Gamma(\Delta) \right)^2}{\tilde{X}^2+\Gamma(\Delta)^2},
\end{equation}
where  $\tilde{X}(\Delta)\equiv\tilde{x}-\tilde{x}_m-\frac{q_F(\tilde{x}_m\Delta)^2}{4\pi(1+q_F^2)}$. 
Here, the Fano parameter $q_F$  and characteristic width  $\Gamma$ are, respectively, given by
\begin{equation}
\begin{aligned}\label{FanoParameters}
   q_F = N \frac{v_0}{\tilde{x}_m} &\quad \text{and}& \Gamma(\Delta)=\frac{(\tilde{x}_m\Delta)^2}{4\pi (1+q_F^{-2})}.
\end{aligned}
\end{equation}
The fact that $\Gamma \sim \Delta^2$ indicates that for any number of impurities ($N$), the Fano profile collapses as $\Delta \rightarrow 0$, implying that the inversion-symmetric configuration ($\theta = \Delta\phi/2$) can host BICs. { Moreover, since the profile is also featured for large values of the dimensionless wavenumbers $ka = mN$ (with corresponding short wavelength $\lambda = a\Delta \phi/N$) the effect is observable over a wide wavelength range, including thus both short and long wavelength regions. Furthermore,}  the dependence of the Fano profile's form on the system's parameters shows the crucial role of interference effects in optical transport. Notably, in the limit where $V \rightarrow 0$, all WGR modes (for any $N$) converge to the isolated impurity-free WGR modes. Consequently, any periodic perturbation in the WGR leads to the appearance of Fano profiles, which indicates that BICs can be formed (even for a single impurity). Conversely, ``turning on'' the periodic impurity potential in the inversion-symmetric configuration suggests that the photonic transport can be directly manipulated by decoupling one of the WGR states at the band edge from the ribbon waveguide. {Finally, we emphasize that Eq.~\eqref{FanoParameters} can be used to directly determine the strength of the impurity potential from the features of the Fano profile.}

\emph{Conclusions and outlook.}  We have studied the optical transport and formation of the BICs in a WGR coupled with impurities and an external optical waveguide mode. In particular, we have derived an exact expression for the transmission probability for a system of $N$ impurities [Eq.~\eqref{transmission amplitude}]. Most importantly, we show that the BICs are formed due to physical conditions that preclude their projection along the external optical waveguide, therefore implying their decoupling from the continuum spectrum and the concomitant localization. In the case of periodically distributed impurities, manipulating the optical transport is particularly prominent when the inversion symmetry is present, allowing for a selective decoupling of resonant states. As such, the proposed setup could be utilized as an efficient impurity detector due to its extreme sensitivity to symmetry. 

When considering the possible applications of BICs and quasi-BICs resonance phenomena, particularly for impurity detection, it is clear that the precise manufacturing of the device will greatly affect its features. Achieving the perfect tangential coupling needed for BIC formation, which is crucial in this respect, may be challenging. This should be achievable when the junction and waveguide widths are similar. We will also explore the impact of more realistic conditions, such as different types of couplings between the waveguide and the WGR. Additionally, these wave phenomena could impact electronic transport at small scales, like in molecules or other complex structures where electron coherence is maintained.

{ As a final remark, we point out that the WGRs have already been employed in experimental studies of  the non-Hermitian systems~\cite{Chen2017}. Therefore, a generalization of our pursuit to non-Hermitian setups should be straightforward by considering local potential with an imaginary part emerging, for instance, from a coupling to an external particle bath, which we plan to address in the near future.}

\emph{Acknowledgment.} This work is supported by the Swedish Research Council Grant No. VR 2019-04735 (V.J.) and Fondecyt (Chile) No. 1230933 (V.J.), DGIIE USM Grant PI-LIR-24-10 (P.A.O.).


\emph{Author Contributions Statement.} M. A. F. performed all the calculations. P.A.O. conceived the project. V. J. and P. A. O. structured and supervised the project.
All the authors contributed to the writing of the manuscript.

\bibliographystyle{naturemag}
\bibliography{Bibliography}
\end{document}